\documentclass[a4paper]{article} 
\usepackage{graphicx}
\usepackage{amsmath,amssymb}

\begin{document}
\thispagestyle{empty} \parskip=12pt \raggedbottom
 
\def\mytoday#1{{ } \ifcase\month \or January\or February\or March\or
  April\or May\or June\or July\or August\or September\or October\or
  November\or December\fi
  \space \number\year}
\noindent
\vspace*{1cm}
\begin{center}
  {\LARGE Absence of evidence for pentaquarks on the lattice} 
\vskip7mm 
 Kieran Holland$^a$ and K.~Jimmy Juge$^b$\footnote{Present address: Department of Physics, Carnegie Mellon University, 5000 Forbes Ave., Pittsburgh, PA 15213.} \\
BGR (Bern-Graz-Regensburg) Collaboration 
\vskip6mm
$^a$
  Department of Physics,
  University of California San Diego, \\
  9500 Gilman Drive,
  La Jolla CA 92093, USA

\vskip1mm

$^b$
  School of Mathematics,
  Trinity College \\
  Dublin 2, Ireland
\vskip1mm 

  \nopagebreak[4]
 
\begin{abstract}

We study the question of whether or not QCD predicts a pentaquark
state $\Theta^+$. We use the improved, fixed point lattice QCD action
which has very little sensitivity to the lattice spacing and also
allows us to reach light quark masses. The analysis was performed on a single volume of size $(1.8\ {\rm fm})^3\times 3.6\ {\rm fm}$ with lattice spacing of $a=0.102$ fm. We use the correlation matrix method to identify the ground and excited states in the isospin 0, negative parity channel. In the quenched approximation where dynamical quark effects are omitted, we do not find any evidence for a pentaquark resonance in QCD.
\end{abstract}
 
\end{center}
\eject

\section{Introduction}

The initial observation \cite{Nakano:2003qx} of a baryonic resonance
$\Theta^+$ with strangeness $S=+1$ generated an enormous amount of
experimental and theoretical activity. As the minimal quark 
content of the $\Theta^+$ is $uudd\bar{s}$, this was the first
evidence for exotic hadronic states with more than three quarks, which have
long been conjectured to exist \cite{Jaffe:1976}. There followed a
number of experimental results  \cite{Stepanyan:2003qr} consistent
with a pentaquark resonance close to $1540$~MeV, roughly $100$~MeV
above the Kaon-Nucleon threshold,  with a surprisingly narrow width,
possibly even as small as a few MeV . However, there have also been
several null experimental results where the $\Theta^+$ resonance has
not been found \cite{Bai:2004gk}. The uncertain experimental status of
the pentaquark is an ideal situation for theoretical calculations to
have an impact, particularly as the parity of the $\Theta^+$ has not
yet been measured (it is expected to have isospin 0). In fact,
previous to the experimental discovery, it was actually predicted
\cite{Diakonov:1997mm} in the chiral soliton model that a narrow
resonance should exist at approximately $1530$~MeV, which was
completely consistent with the first observations. There have since
been a number of attempts \cite{Jaffe:2003sg} to model the behavior of
the pentaquark, in particular its very narrow width, for example using
quark models, Kaon-Nucleon bound states and QCD sum rules.  

The only known method to derive hadronic properties {\it ab initio}
from a theory of strongly-interacting quarks and gluons is lattice QCD. Not
surprisingly, there have been a number of lattice studies to see
if the $\Theta^+$ can be predicted from QCD. The first attempts used
standard hadron spectroscopy methods to explore pentaquark states in a
number of isospin and parity channels, to determine for which quantum
numbers could a pentaquark state be consistent with experiment
\cite{Csikor:2003ng, Sasaki:2003gi, Chiu:2004gg}. It was
consistently found that the lightest isospin 0 state is below the lightest
isospin 1 state, but there was disagreement on the parity assignment
of a possible $\Theta^+$ resonance. There were also studies which did
not find a pentaquark resonance, only scattering states of
weakly-interacting Kaons and Nucleons \cite{Mathur:2004jr,Ishii:2004qe}. More
recent studies which try to disentangle a possible pentaquark from
Kaon-Nucleon states indicate that a pentaquark resonance is not
seen \cite{Lasscock:2005tt,Csikor:2005xb}. However consensus has not
yet been reached, as other work
\cite{Chiu:2004gg, Alexandrou:2005gc,Takahashi:2005uk} continues to favor a
single-particle resonance consistent with the experimental $\Theta^+$. 
The search for spin-$3/2$ pentaquarks on the lattice have also been
conducted more recently by a couple of groups (\cite{Ishii:2005vc,Lasscock:2005kx}); however, also without agreement on 
their conclusions.

The purpose of this short paper is to try to add to the level of
agreement among lattice QCD calculations. There are countless ways QCD
can be regulated using the lattice as a space-time cutoff, but all
possible formulations should agree in their physical predictions when
the lattice spacing vanishes and the continuum limit is reached. We
use a particular improved lattice QCD action, designed so that
physical continuum results can be extracted even on relatively coarse
lattices. Because our action is also chirally symmetric at vanishing
quark mass (up to small parametrization errors), this allows
us to reach light quark masses, which is not possible in many
lattice formulations. We use the correlation matrix method to separate
a possible pentaquark resonance from an interacting multiparticle
system. As in the other studies, we use the quenched
approximation where dynamical quark effects are omitted, which is an
enormous saving in computational difficulty. Although without
justification, this approximation empirically gives qualitatively and
even quantitatively the correct picture for the hadronic states whose
width is narrow. The expectation is that this will also hold for any
possible pentaquark states, if it is indeed as narrow as some
experimental results claim. 

The paper is organised as follows: in Section 2 we give a short
description of the lattice action and the details of the numerical
simulation. In Section 3 we discuss the correlation matrix analysis to
separate ground and excited states in various channels. In Section 4 we
present and discuss our numerical results, followed by our conclusions.

\section{Simulation details}

The lattice QCD action we use is based on the real space renormalization
group \cite{Hasenfratz:1993sp}. The fixed point action is a close
approximation to the exact renormalization group trajectory, so the
theory has only a weak dependence on the lattice spacing and continuum
results can be extracted even on relatively coarse lattices.
If the continuum limit is reached earlier, this could bring a large reduction in the
computational time required to work at a fixed physical volume. Fixed
point actions have been constructed for 2- and 4-dimensional systems
and in general have shown the expected good behavior
\cite{Wiese:1993cb}. In addition, the exact fixed point Dirac operator
is a solution to the Ginsparg-Wilson relation \cite{Ginsparg:1981bj},
giving exact chiral symmetry at non-zero lattice spacing
\cite{Hasenfratz:1997ft}. The fixed point action is one of the many
possible lattice fermion formulations \cite{Narayanan:1994gw} that
have all the desirable chiral properties of the continuum theory, such
as massless pions at vanishing quark mass. The action we use in the
simulations is close to an exact Ginsparg-Wilson solution, hence the
chiral properties are almost completely intact.  

The fixed point QCD action has been tested in a number of quenched QCD
studies \cite{Hasenfratz:2000xz}, including hadron spectroscopy,
$\pi$-$\pi$ scattering lengths, the chiral condensate, glueball
masses, the topological susceptibility, unphysical zero-mode effects
in quenched QCD, and properties of the Dirac operator
eigenvalues. Large-scale hadron spectroscopy studies have shown the
improved behavior of the action
\cite{Gattringer:2003qx,Hasenfratz:2004qk}. Hadron masses calculated
at lattice spacings of $0.153$ and $0.102$~fm have shown only a weak
dependence on the lattice cutoff and are closer to
continuum-extrapolated values than using the Wilson action at
$0.05$~fm or the improved staggered action at $0.09$~fm. Pion masses
as light as $220$~MeV could be reached without any indication of
exceptional configurations (i.e.~gauge field configurations for which
the Dirac operator has unphysical accidentally small eigenvalues due
to large fluctuations), which limit the lightest possible quark mass
that can be reached. This is an indication of the good chiral behavior
of the lattice action. 

In this paper we study an ensemble of 176 gauge configurations\footnote{Four random configurations were irretrievable from the backup tapes making the total number of configurations 176 as opposed to the 180 which were used in Ref.\cite{Hasenfratz:2004qk}. The conclusions are not affected by this loss in any way.} of
lattice size $18^3 \times 36$ which have already been generated for
standard hadron spectroscopy \cite{Hasenfratz:2004qk}. From the Sommer
parameter $r_0 = 0.49$~fm, the lattice spacing is found to be
$a(r_0)=0.102$~fm, where the systematic error is at the \% level and
the statistical error is less than $1\%$. The gauge configurations are
separated by 500 alternating Metropolis and pseudo-over-relaxation
sweeps. The gauge configurations are fixed to the Coulomb gauge and
Gaussian-smeared sources of width  
$\sigma/a\sim2.3$ were used for the quark propagators. The ensemble covers a range of quarks masses, reaching
pseudoscalar meson masses as light as $m_{\it PS}=300$~MeV.

\section{Analysis}

The usual goal in standard hadron spectroscopy studies is to extract
particle masses for the various possible quantum numbers. The typical method
used is to calculate the correlation functions in Euclidean time of two-
and three-quark operators with the desired quantum numbers. For example,
the pion correlation function is 
$C_{\pi}(t) = \langle \Omega | {\cal O}(t) {\cal O}^{\dagger}(0)
| \Omega \rangle$, where the pion operator is ${\cal O} = \bar{\psi}
\gamma_5 \psi$. For large Euclidean time, the correlation function is
dominated by the lowest energy state
\begin{equation}
C(t) \propto | \langle \Omega | {\cal O} | 0 \rangle |^2 \exp(-E_0 t).
\end{equation}
If the correlation function can be measured for sufficiently large time
separations, the higher energy states are suppressed and this
is a safe method to extract the ground state energy. This can be seen in the
effective mass 
\begin{equation}
m_{\rm eff}(t) = -\ln[ C(t+1)/C(t) ],
\end{equation}
which develops a plateau when the higher states give only a negligible
contribution to the correlation function. However, if the energy
separations are small, the higher energy states die out slowly and
contaminate the lowest state. In addition, if the overlap of the
operator with the ground state is accidentally small, it might be
necessary to measure the correlation function for very large time
separations to suppress the higher states. In these situations, the time
separation required for the effective mass to develop a plateau might
be unfeasibly large. In principle, one can extract the energy of the
excited states by including their contribution to the correlation
function. However, in practice, this requires measuring $C(t)$ to
greater accuracy than is normally possible.

A more reliable method to extract the energy states is to form a matrix of
correlation functions $C_{ij}$ of operators with the same desired quantum
numbers \cite{Michael:1985ne,Luscher:1990ck}. For an $n \times n$
correlation matrix, the solution of the generalized eigenvalue problem  
\begin{equation}
\sum_j C_{ij}(1) b_j^{(n)} = \lambda_n \sum_j C_{ij}(0) b_j^{(n)}
\end{equation}
between timeslices $t=0$ and $t=1$ is used to construct the correlators 
\begin{equation}
\tilde{C}_n(t) = \sum_{i,j} b_i^{(n)\dagger} C_{ij}(t) b_j^{(n)} 
\end{equation}
whose overlap with the $n$ lowest energy states is optimal. The
contamination of the ground state by higher states is minimized and
the excited states themselves can also be extracted if the correlation
matrix is measured to sufficient accuracy. The correlation matrix
analysis is more reliable for extracting the ground state and it is
necessary for getting higher energy states for multiparticle
systems. If the $\Theta^+$ state does exist, its quantum numbers will be
the same as a tower of Kaon-Nucleon interacting states. If the
pentaquark is not the lightest state in that channel, at the very
least the two lightest states must be extracted to positively identify
the $\Theta^+$. 

To look for a possible pentaquark state $\Theta^+$, some five-quark
operator with the desired quantum numbers must be chosen. To construct
a correlation matrix, more than one such operator is required. Without
much guidance, it is difficult to tell which will have the best overlap with
the $\Theta^+$ wave function. A number of operators have been tried,
including ones based on a Kaon-Nucleon-type
construction or using the diquark-diquark-antiquark structure proposed
by Wilczek and Jaffe. Many possible five-quark operators can be
related to one another by Fierz transformations and all of them must
have some overlap with Kaon-Nucleon states. Even with the experience
of the previous lattice QCD studies, it is not clear what the best
choice is. To allow us to make a direct comparison with other work, we
use the operator originally suggested in \cite{Zhu:2003ba} and first used in a lattice simulation in \cite{Csikor:2003ng}
\begin{equation}
\Theta = \epsilon^{abc} [u^T_a C \gamma_5 d_b]\{ u_e \bar{s}_e i
\gamma_5 d_c \mp (u \leftrightarrow d) \},
\end{equation}
as well as the color-rearranged combination
\begin{equation}
KN = \epsilon^{abc} [u^T_a C \gamma_5 d_b]\{ u_c \bar{s}_e i
\gamma_5 d_e \mp (u \leftrightarrow d) \},
\end{equation}
where $C=\gamma_2 \gamma_4$ is the charge-conjugation matrix and the
minus and plus signs correspond to the isospin 0 and 1 combinations
respectively. These operators have intrinsic negative parity, however,
it is well-known that the correlation functions constructed from these
operators receive contributions from negative and positive parity
states. To project a particular parity state, an additional factor $(1
\pm \gamma_4)$ is included in the correlation function.

The quark propagators were generated from Gaussian-smeared
sources. Using a smeared instead of point source has previously been
shown to accelerate the decay of excited states in correlation
functions and to generate effective masses whose plateaux extend to
smaller time separation, allowing a more accurate determination of the
effective mass. Smearing both the source and the sink in the
correlation function has not in general shown much additional
improvement. In our study, we use smeared sources and point
sinks. This asymmetry means the correlation matrix is not Hermitian
and so left and right eigenvectors must be determined in the
generalized eigenvalue problem to construct the optimal correlators: 
\begin{eqnarray}
&&\sum_j C_{ij}(1) b_j^{(n)} = \lambda_n \sum_j C_{ij}(0) b_j^{(n)},
\nonumber \\
&&\sum_i a_i^{(n)\dagger} C_{ij}(1) = \lambda_n \sum_i a_i^{(n)\dagger} C_{ij}(0),
\nonumber \\
&&\tilde{C}_n(t) = \sum_{i,j} a_i^{(n)\dagger} C_{ij}(t) b_j^{(n)}.
\end{eqnarray}

\section{Numerical results}

Previous lattice studies have examined a variety of isospin and parity
channels to find a state compatible with the observed $\Theta^+$. With the
exception of one study, a possible signal of a genuine pentaquark
state has been found only in the isospin 0 negative parity
channel. For the purpose of this short paper, we will concentrate on
these quantum numbers. 

For this ensemble of gauge configurations, quark propagators were
calculated for 10 quark masses, with pseudoscalar meson masses ranging
from $m_{\it PS}=300$~MeV to 1390~MeV. Although one is ultimately
interested in reaching the physical quark masses, it is also
instructive to study the mass dependence of hadronic states. To
calculate a correlation function of five-quark operators, the number
of quark propagator contractions is much greater than for two- or
three-quark operators. Hence we consider only some of the possible
combinations of up-down and strange masses, $m_{ud}$ and $m_s$, for
this ensemble. In quenched QCD, where dynamical quark effects are
omitted, gauge configurations with topological zero modes of the Dirac
operator are not suppressed in the ensemble when the quark masses
become small. The zero modes can give large unphysical 
${\cal O}(1/m_q^n)$ contributions to correlation functions, which are
only suppressed as the volume becomes large. This effect and possible
solutions have been studied extensively for two- and three-quark correlation
functions, but not yet for the pentaquark case. Since our physical
volume is relatively small, roughly 1.8~fm, this might prevent us from
extracting useful information from the lightest quark masses.

\begin{figure}[t]
\begin{center}
\includegraphics[width=1.1\textwidth,height=0.95\textwidth]
{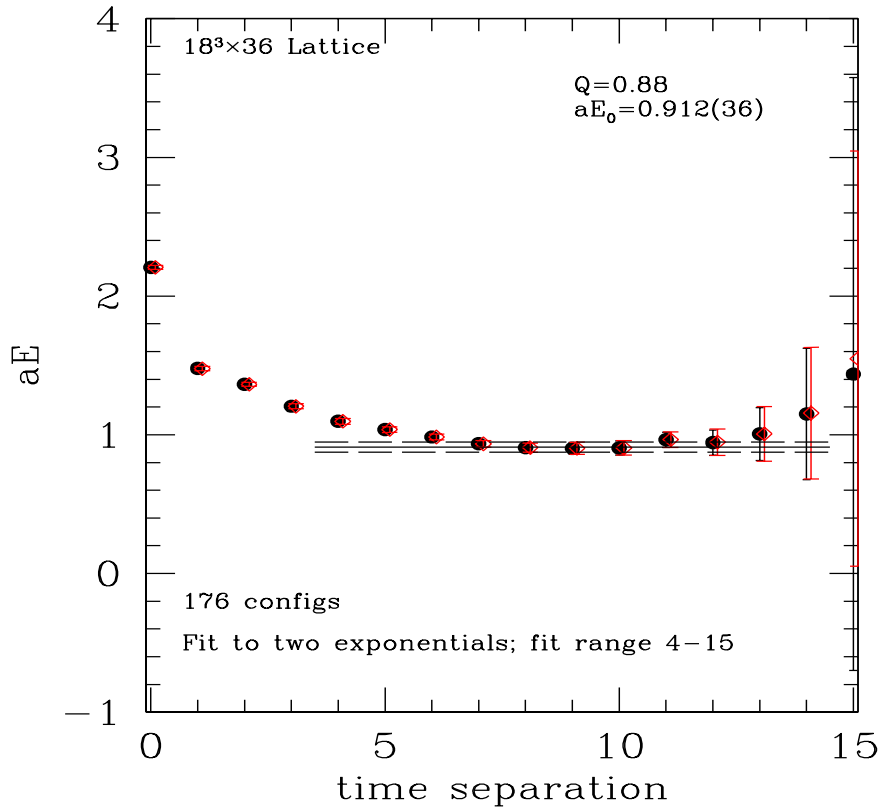}
\end{center}
\vspace{-5mm}
\caption{{}\label{fig:effmass0} The effective mass of the optimized
  pentaquark ground state correlator. The optimal effective mass as defined in Ref.~\cite{Luscher:1990ck} is also shown with diamond symbols. }
\end{figure}

\begin{figure}[t]
\begin{center}
\includegraphics[width=1.1\textwidth,height=0.95\textwidth]
{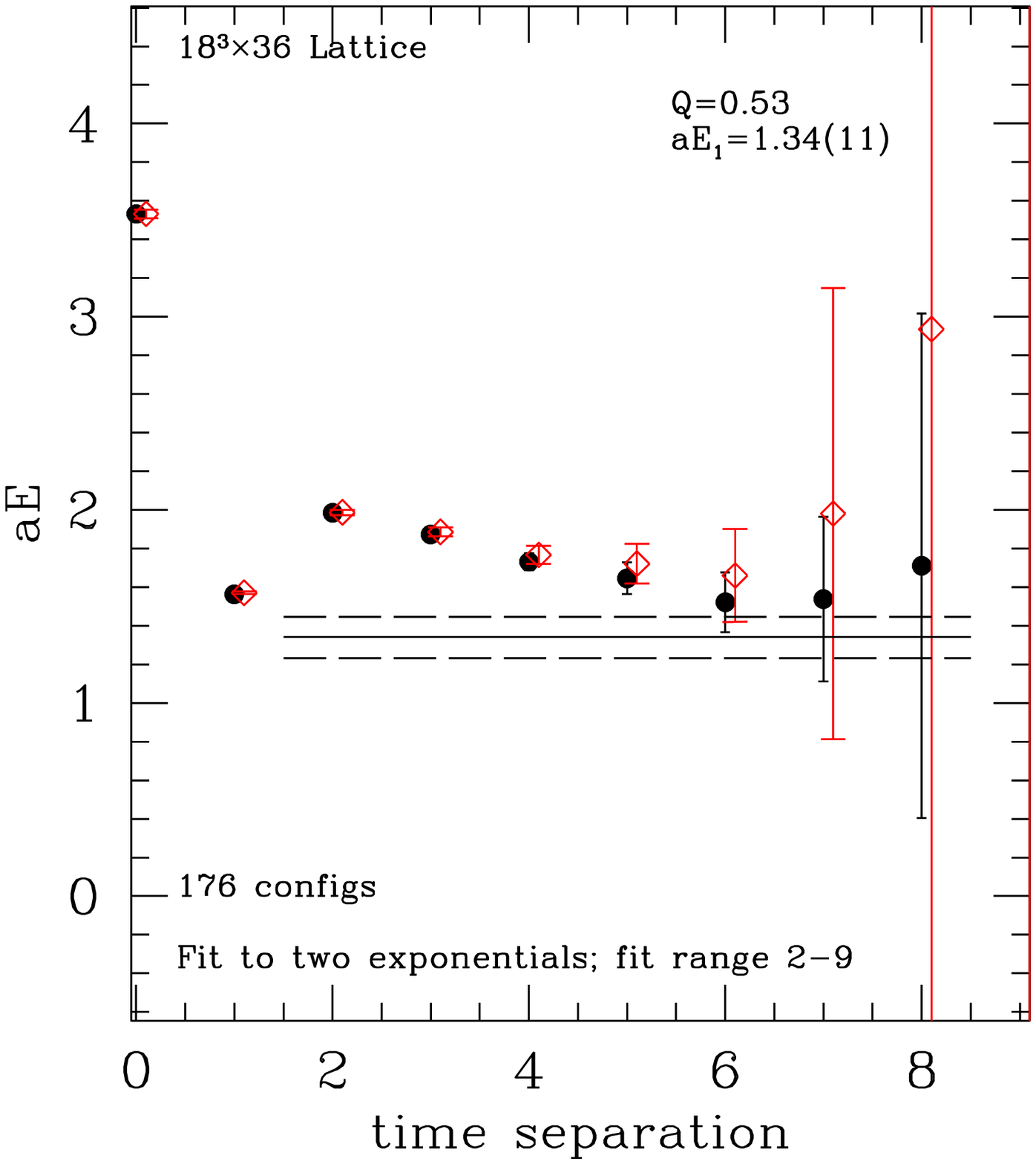}
\end{center}
\vspace{-5mm}
\caption{{}\label{fig:effmass1} The effective mass of the optimized pentaquark first excited state correlator. The optimal effective mass as defined in Ref.~\cite{Luscher:1990ck} is also shown with diamond symbols.}
\end{figure}

\begin{figure}[t]
\begin{center}
\includegraphics[width=1.1\textwidth,height=0.95\textwidth]
{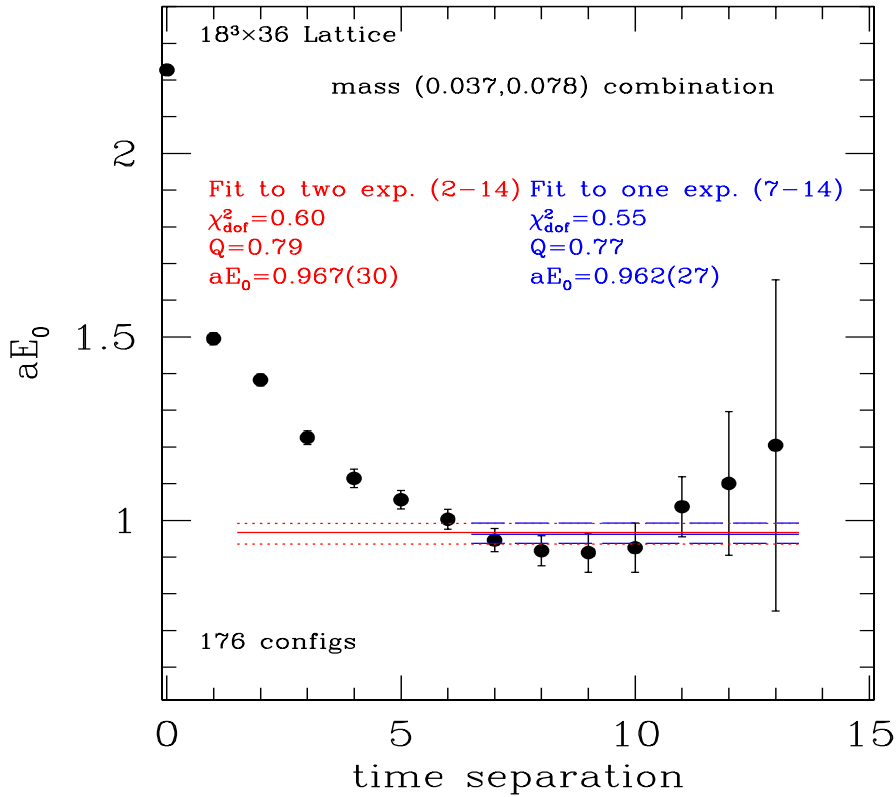}
\end{center}
\vspace{-5mm}
\caption{{}\label{fig:agreement} Single- and two-exponential fits to
  the ground state diagonal component of the optimized correlation
  matrix. For the single-exponential fit, the fit could only be done
  starting at large time separations.}
\end{figure}

\begin{figure}[t]
\begin{center}
\includegraphics[width=1.05\textwidth,height=0.9\textwidth]
{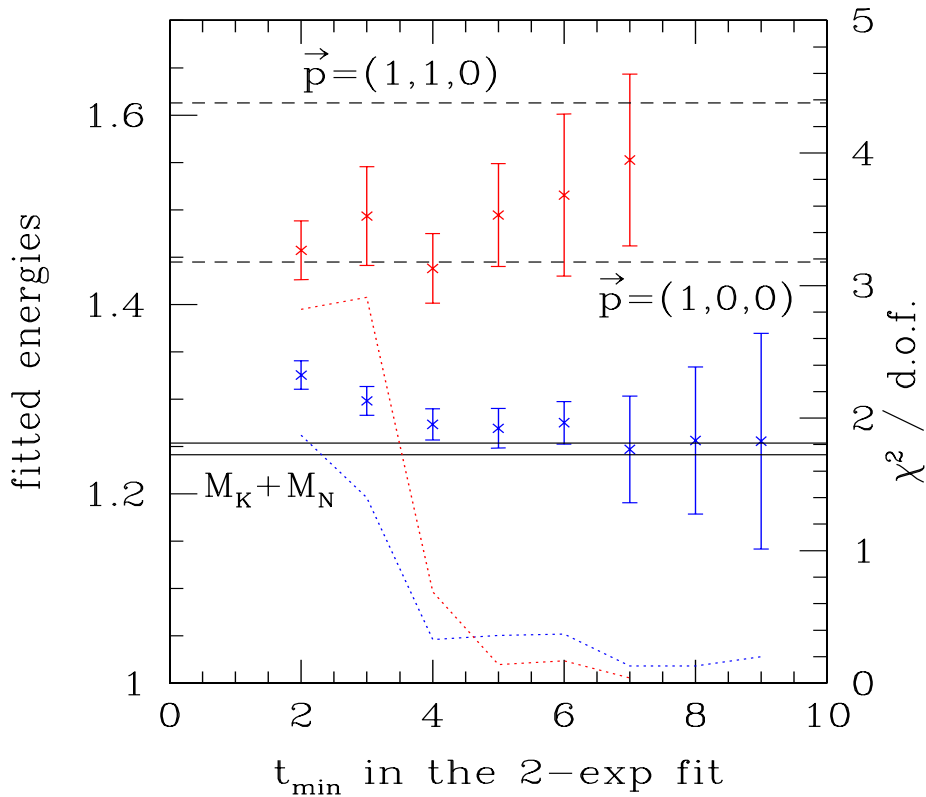}
\end{center}
\vspace{-5mm}
\caption{{}\label{fig:fits} The fitted energies for the pentaquark
  ground and first excited states and the quality of the fits as
  a function of $t_{\it min}$, the smallest time separation included
  in the fit range. The horizontal solid lines are the Kaon-Nucleon threshold
  (including the error) and the horizontal dashed lines are the energy
  levels of Kaon-Nucleon scattering states.}
\end{figure}

We calculate the $2 \times 2$ correlation matrix of the operators
$\Theta$ and $KN$ and construct the optimal correlators
$\tilde{C}_{ij}(t)$. Near the region where a plateau in the effective
mass is observed, we fit the diagonal entries of the optimal correlators to a two-exponential ansatz
\begin{eqnarray}
&& \tilde{C}_{00}(t) = A_0 e^{-E_0t}(1 + B_0 e^{-\Delta_0t}) \nonumber \\
&& \tilde{C}_{11}(t) = A_1 e^{-E_1t}(1 + B_1 e^{-\Delta_1t}),
\end{eqnarray}
giving us the ground and first excited energies $E_0$ and $E_1$. The
best fit values were obtained using the standard correlated $\chi^2$
method using the measured covariance to estimate the covariance
matrix. The errors were obtained by a 1024-point bootstrap sampling
procedure. The different quark mass combinations were fitted
independently and so correlations between the different quark mass
values were not taken into account. We should note that there are also
backward-propagating contributions to the correlators, but these are
suppressed by a factor of $\sim 100$ in the region where the
two-exponential fits are performed. 
In Figures~\ref{fig:effmass0} and \ref{fig:effmass1} we show typical
effective mass plots and the corresponding two-exponential fits for
the pentaquark ground and first excited state. A
single-exponential fit can also be used, but only over a shorter time
range, which limits the precision of the fitted parameters. The
single- and two-exponential fitting procedures give completely
consistent results, as shown in Figure~\ref{fig:agreement}. To
demonstrate the stability of the two-exponential fits, we show in
Figure~\ref{fig:fits} the fitted energies for a particular quark mass
combination as a function of the minimum timeslice included in the
fits. The dotted lines are the $\chi^2$ per degree of freedom from the
fits and the horizontal solid and dashed lines indicate the Kaon-Nucleon
threshold and lowest non-zero momentum scattering states. The fitting
procedure clearly works well.

\begin{table}[t]
\begin{center}
\renewcommand{\arraystretch}{1.2} 
\begin{tabular}{rrrrrrrr} 
\hline \hline
$am_{ud}$ & $am_s$ & $t_{\it min}$ & $t_{\it max}$ & 
$\chi^2_{\it dof}$ & $Q$ & $E_0$ & $E_1$ \\
\hline \hline
0.029 & 0.078 & 3 & 12 & 0.84 & 0.54 & 0.889(87) & - \\
      &       & 2 & 7  & 1.58 & 0.21 &        - & 1.31(20) \\ 
\hline
0.032 & 0.078 & 3 & 13 & 0.74 & 0.64 & 0.903(67) & - \\
      &       & 2 & 7  & 3.02 & 0.05 &         - & 1.35(15) \\
\hline
0.037 & 0.078 & 2 & 14 & 0.60 & 0.79 & 0.967(30) & - \\
      &       & 2 & 8  & 2.57 & 0.05 &         - & 1.37(12) \\
\hline
0.045 & 0.078 & 4 & 15 & 0.45 & 0.89 & 0.981(35) & - \\
      &       & 2 & 9  & 1.90 & 0.11 &         - & 1.360(91) \\
\hline
0.037 & 0.037 & 2 & 11 & 0.60 & 0.61 & 0.844(45) & - \\
      &       & 2 & 7  & 0.57 & 0.57 &         - & 1.40(21) \\
\hline
0.045 & 0.045 & 4 & 15 & 0.47 & 0.88 & 0.912(36) & - \\
      &       & 2 & 9  & 0.78 & 0.53 &         - & 1.34(11) \\
\hline
0.058 & 0.058 & 5 & 16 & 0.50 & 0.86 & 1.027(28) & - \\
      &       & 2 & 11 & 1.74 & 0.11 &         - & 1.324(60) \\
\hline
0.078 & 0.078 & 4 & 14 & 0.20 & 0.99 & 1.140(20) & - \\
      &       & 5 & 11 & 0.43 & 0.73 &         - & 1.388(75) \\
\hline
0.100 & 0.100 & 4 & 14 & 0.33 & 0.94 & 1.273(17) & - \\
      &       & 4 & 11 & 0.69 & 0.60 &         - & 1.438(36) \\
\hline
0.100 & 0.140 & 4 & 14 & 0.47 & 0.86 & 1.329(16) & - \\
      &       & 4 & 11 & 0.71 & 0.59 &         - & 1.527(42) \\
\hline
0.140 & 0.140 & 5 & 14 & 0.64 & 0.70 & 1.498(18) & - \\
      &       & 4 & 12 & 0.41 & 0.84 &         - & 1.646(34) \\
\hline
0.180 & 0.180 & 4 & 14 & 0.64 & 0.72 & 1.716(13) & - \\
      &       & 4 & 12 & 0.10 & 0.99 &         - & 1.863(35) \\
\hline
0.240 & 0.240 & 4 & 15 & 1.08 & 0.37 & 2.011(16) & - \\
      &       & 4 & 13 & 1.12 & 0.35 &         - & 2.120(43) \\ 
\hline
\end{tabular}
\end{center}
\caption{{} The fitted energy values for the ground and first excited
  states. The two states were fitted separately to two exponentials
  after the diagonalization of the $2 \times 2$ correlation matrix.
 \label{table:1}}
\end{table} 

\begin{table}[t]
\begin{center}
\renewcommand{\arraystretch}{1.2} 
\begin{tabular}{rrrrrrr} 
\hline \hline
$am_{ud}$ & $am_s$ & $t_{\it min}$ & $t_{\it max}$ & 
$\chi^2_{\it dof}$ & $Q$ & $aM_{\it PS}$ \\
\hline \hline
0.029 & 0.078 & 5 & 12 & 1.36 & 0.23 & 0.2768(15) \\
0.032 & 0.078 & 5 & 12 & 1.35 & 0.23 & 0.2817(15) \\
0.037 & 0.078 & 6 & 11 & 1.03 & 0.39 & 0.2884(17) \\
0.045 & 0.078 & 5 & 12 & 1.41 & 0.21 & 0.3031(14) \\
0.100 & 0.140 & 5 & 10 & 1.16 & 0.33 & 0.4656(15) \\
\hline
\end{tabular}
\end{center}
\caption{{} The pseudoscalar meson mass for non-degenerate quark
  masses.
 \label{table:2}}
\end{table}

The meson and nucleon masses for degenerate $m_s$ and $m_{ud}$ quarks
have previously been calculated for this ensemble and are tabulated in
\cite{Hasenfratz:2004qk}. In Tables~\ref{table:1} and \ref{table:2} we
give the results for the five-quark ground and first excited state
energies, and the pseudoscalar meson mass for non-degenerate
quarks. Most of the data are for equal up-down and strange quark
masses. The physical value of the strange quark mass is very close to
$am_s=0.078$, so we also consider non-degenerate quark masses with the
strange quark mass fixed at this value. 

\begin{figure}[t]
\begin{center}
\includegraphics[width=1.05\textwidth,height=0.85\textwidth]
{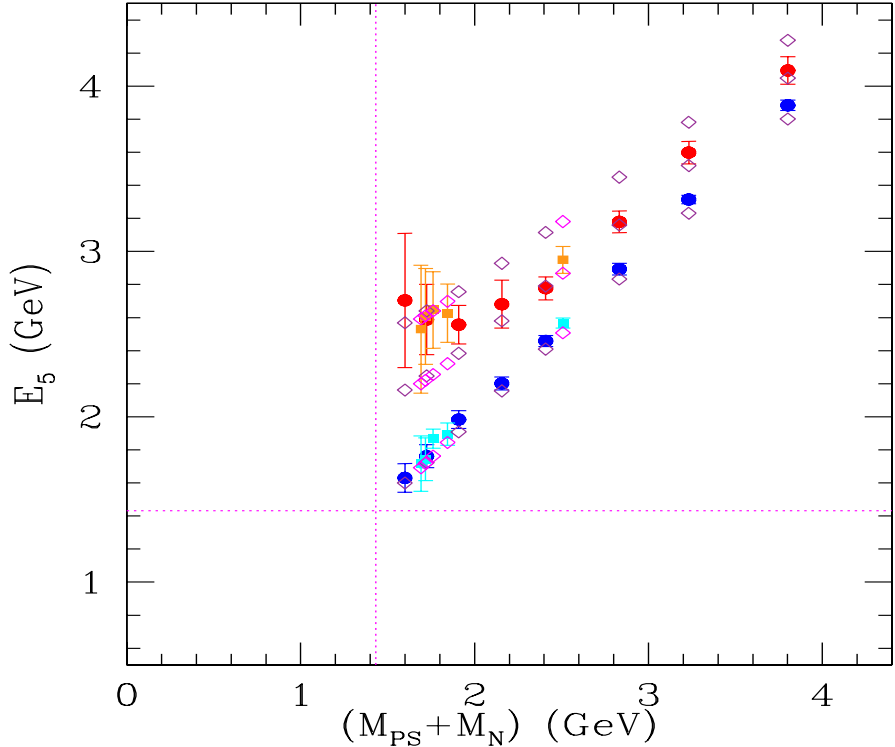}
\end{center}
\vspace{-5mm}
\caption{{}\label{fig:summary} The fitted pentaquark ground and first
  excited state energies as a function of the sum of the nucleon and
  pseudoscalar meson mass. The degenerate quark mass data are the dark
  blue and red circles, the non-degenerate quark mass data the light
  blue and orange squares. The hollow symbols are the three lowest
  Kaon-Nucleon scattering state energies. The dashed lines indicate
  the physical values of $M_K+M_N$.}
\end{figure}

Previous lattice QCD and experimental studies have shown that the
Kaon-Nucleon interaction in the isospin 0 s-wave channel is
weak \cite{Fukugita:1994ve,Hyslop:1992cs}. Therefore the energy of a
Kaon-Nucleon scattering state with equal back-to-back momentum is
closely approximated by  
\begin{equation}
E_{\it KN}(p) = \sqrt{M_K^2 + |\vec{p}|^2} + \sqrt{M_N^2 + |\vec{p}|^2}, 
\end{equation}
where the momentum on a finite lattice is quantized $\vec{p}=2 \pi \vec{n}/L$.  
In Figure~\ref{fig:summary} we plot the five-quark ground and first
excited state energies and the three lowest Kaon-Nucleon scattering
state energies as a function of the sum of the nucleon and
pseudoscalar meson masses. Over a large range of quark masses, we see
that the five-quark energies are in very good agreement with the
weakly-interacting scattering states. As the quark masses are reduced,
the five-quark excited state is more difficult to determine, but the
ground state is still accurately measured. With $m_s$ close to the
physical value and $m_{ud} < m_s$, the five-quark ground state agrees
well with the Kaon-Nucleon ground state but we cannot make a strong
statement about the excited state. 
The most important observation is that, for
heavier quarks where our results are most accurate, we do not see any
indication of a pentaquark resonance state, which should lie between
the scattering state energy levels. The picture for light quarks in
consistent with this, where the first excited state lies between the
first and second excited Kaon-Nucleon scattering states, with much
larger statistical errors. In principle, one would like to
extrapolate the results to the physical quark masses. However, as we
mentioned previously, quenched QCD correlation functions suffer from
unphysical large contributions from topological zero modes at small
quark mass. This complication might obstruct taking the chiral limit
on this lattice volume. In addition, one would like to repeat this
calculation for a range of lattice spacings to extrapolate to the
continuum limit. However other hadronic states have shown only a very
weak dependence on the lattice cutoff, so we expect this will not
change our results significantly. 

Let us compare our results with other studies. The first attempts\footnote{One of the referees of this paper has requested to emphasize at this point that the quark-exchange diagram between diquark pairs have been omitted in Ref.~\cite{Sasaki:2003gi}.}
\cite{Csikor:2003ng, Sasaki:2003gi} at pentaquark spectroscopy found
an energy state slightly above the Kaon-Nucleon threshold in the
isospin 0 negative parity channel, which looked consistent with the
experimental $\Theta^+$. Further studies using Bayesian techniques to
extract ground and excited states found energies consistent with K-N
scattering only \cite{Mathur:2004jr}. Scattering states can also be
distinguished from resonances via the volume-dependence of the
energies, which can also be raised with hybrid boundary conditions
\cite{Ishii:2004qe}. The amplitudes $W_n$ of the energy states
contributing to the correlation functions also have a distinctive
volume-dependence for two-particle scattering. The correlation matrix
method has also been used previously to extract the pentaquark ground
and excited state energies 
\cite{Lasscock:2005tt, Csikor:2005xb, Takahashi:2005uk}. 
There is growing evidence that the isospin 0
negative parity channel, the one thought most likely to describe the
experimentally observed $\Theta^+$, only has only weakly-interacting
scattering states and no resonance. A stronger claim is made in
\cite{Lasscock:2005tt}, that a genuine pentaquark resonance should
become a bound state at heavy quark mass, which is clearly not seen in
any study. However, the picture is not completely consistent. In
\cite{Takahashi:2005uk} the pentaquark first excited energy state
determined from the correlation matrix does not have the expected
volume-dependence of two-particle scattering in small volumes. In
\cite{Takahashi:2005uk, Alexandrou:2005gc} the amplitudes $W_n$ for
the excited state appear to be independent of volume, signalling a
resonance rather than a two-particle state, in contradiction to the
findings in \cite{Mathur:2004jr}. There is also the competing claim
\cite{Chiu:2004gg} that the $\Theta^+$ is actually in the isospin 0
positive parity channel.  

\section{Conclusions}

Our conclusions are relatively straightforward. The correlation matrix
technique is designed to extract a number of energy levels in a given
channel. Previous evidence suggests that the physical $\Theta^+$, if it
exists, is in the isospin 0 negative parity channel, above the
Kaon-Nucleon threshold. In that case, the minimum requirement from a
lattice study is to identify the two lowest energy states. Doing this,
we can accurately determine the five-quark energies for a range of
quark masses. We find that the ground and first excited states are
completely consistent with weakly-interacting Kaon-Nucleon scattering
states, with no indication of a pentaquark resonance which should lie
between them. Although we work at a finite lattice spacing, previous
work has shown that this lattice action reproduces extremely well continuum
properties of hadronic states, so we do not expect our observation to
change significantly in the continuum limit.

As the title of this paper suggests, absence of evidence is not
evidence of absence and there are many caveats one can include. This
and all other lattice QCD pentaquark studies have ignored the effect
of dynamical fermions, which is an unjustified approximation. The most
accurate results ruling out a pentaquark resonance are at unphysically
heavy quark masses, requiring an extrapolation to the physical
values. We work at a finite lattice spacing and do not extrapolate to the 
continuum. However we do not expect these effects to qualitatively change our
conclusions. We believe a more serious issue is our choice of operators in
the correlation matrix, which may have little overlap with a genuine $\Theta^+$
state, if it exists. We cannot rule out this possibility, and there may even be
other reasons why we do not see a pentaquark resonance. However we can only
conclude that we find no evidence that QCD predicts the $\Theta^+$ state.  

\section{Acknowledgements}

We wish to thank Peter Hasenfratz, Julius Kuti, and Aneesh Manohar
for invaluable discussions and support and Shoichi Sasaki for comments 
and discussions on our manuscript. We wish to thank
the Swiss Center for Scientific computing in Manno, where the
numerical simulations were performed. This work was supported by the
IITAC PRTLI initiative, by the Department of Energy under the grant
DOE-FG03-97ER40546 and by the National Science Foundation under Grant
No.PHY99-07949. KH wishes to thank the KITP at the University of
California Santa Barbara and the ITP at the University of Bern, where
this paper was partly written.  

\clearpage


\eject

\end{document}